\documentclass[aps,prb,twocolumn]{revtex4}
\usepackage{bm,color,amsmath,amssymb,mathrsfs,latexsym,graphicx,psfrag}

\newcommand{\bk}{{\mathbf k}}
\newcommand{\bq}{{\mathbf q}}

\newcommand{\be}{\begin{equation}}
\newcommand{\ee}{\end{equation}}

\def\bea{\begin{eqnarray}}
\def\eea{\end{eqnarray}}

\begin{document}
\title{Topological nodal line semimetals with and without spin-orbital coupling}
\author{Chen Fang$^{1\ast}$, Yige Chen$^{2,3}$, Hae-Young Kee$^{2,3}$ and Liang Fu$^1$}
\affiliation{$^1$Department of physics, Massachusetts Institute of Technology, Cambridge, MA 02139}
\affiliation{$^2$Department of Physics, University of Toronto, Toronto, Ontario, M5S1A7, Canada}
\affiliation{$^3$Canadian Institute for Advanced Research/Quantum Materials Program, Toronto, Ontario MSG 1Z8, Canada}
\date{\today}
\begin{abstract}
We theoretically study three-dimensional topological semimetals (TSMs) with nodal lines protected by crystalline symmetries. Compared with TSMs with point nodes, e.g., Weyl semimetals and Dirac semimetals, where the conduction and the valence bands touch at discrete points, in these new TSMs the two bands cross at closed lines in the Brillouin zone. We propose two new classes of symmetry protected nodal lines in the absence and in the presence of spin-orbital coupling (SOC), respectively. In the former, we discuss nodal lines that are protected by the combination of inversion symmetry and time-reversal symmetry; yet unlike any previously studied nodal lines in the same symmetry class, each nodal line has a $Z_2$ monopole charge and can only be created (annihilated) in pairs. In the second class, with SOC, we show that a nonsymmorphic symmetry (screw axis) protects a four-band crossing nodal line in systems having both inversion and time-reversal symmetries.
\end{abstract}
\maketitle

The study of topological semimetals has recently drawn much attention from both the theoretical and the experimental communities. Topological semimetals (TSMs) are systems where the conduction and the valence bands have robust crossing points in $k$-space, compared with normal metals where the two bands have a direct gap at each $\bk$. Compared with normal metals, the Fermi surface (FS) of an ideal TSM has a reduced dimension: in two dimensions, a normal metal has one-dimensional (1D) FS while a TSM has zero-dimensional (0D) FS; and in three dimensions, a normal metal has two-dimensional (2D) FS while a TSM has 1D or 0D FS. More importantly, the states near the FS are characterized by a nontrivial topological number. These unique features of FS in TSMs give rise to exotic properties such as the existence of Fermi arcs on the surface\cite{Wan2011} and the chiral anomaly in the bulk\cite{Hosur2012,Son2013}. In three dimensions, Weyl semimetals\cite{Wan2011} and Dirac semimetals\cite{Young2012} have been intensively studied both theoretically\cite{Burkov2011,Burkov2011a,Hosur2012,Fang2012,Liu2013,Lu2013,Weng2015,Huang2015,Wang2012,Zeng2015} and experimentally\cite{Lu2015,Xu2015a,Lv2015,Zhang2015,Huang2015a,Liu2014,Liu2014a,Neupane2014,He2014,Jeon2014,Xu2015,Xiong2015}. In Weyl semimetals, the two bands cross at an even number of discrete points in the BZ, around which the bands are non-degenerate and disperse linearly in all three directions. In Dirac semimetals, both the conduction and the valence bands are twofold degenerate and cross each other at an odd or even number of points. Both systems belong to the class of topological nodal-point semimetals (TPSM).

In three dimensions, there is another class of TSMs where the conduction and the valence bands cross each other at closed lines instead of discrete points, i.e., nodal lines\cite{Burkov2011b,Carter2012,Chiu2014,Phillips2014,Chen2015,Mullen2015,Weng2014,Xie2015,Zeng2015,Kim2015,Yu2015,Rhim2015,Chiu2015,Bian2015}. These nodal-line semimetals (TLSM) are in the midway between TPSMs and normal metals: (i) at exact half-filling, the FS is 0D, 1D and 2D in TPSMs, TLSMs and normal metals and (ii) the density of state scales as $\rho_0\propto{(E-E_f)^2}$, $\rho_0\propto{|E-E_f|}$ and a constant in TPSMs, TLSMs and normal metals, a fact from which one expects distinct electron correlation effects in the three classes. For example, the screening effect in these new metallic states have been discussed\cite{Huh2015}. Very recently, many theoretical proposals of materials for realizing TLSM have emerged, including the realization in graphene networks\cite{Weng2014}, in Ca$_3$P$_2$\cite{Xie2015}, in LaN\cite{Zeng2015}, and in Cu$_3$(Pd,Zn)N\cite{Kim2015,Yu2015}. In all these works, and also in earlier theoretical model studies\cite{Burkov2011b,Chiu2014,Phillips2014}, the nodal line has the following properties: (i) unlike a Weyl node, a single line node can shrink to a point and vanish by continuously tuning the Hamiltonian\cite{Kim2015} and (ii) its stability requires the absence of spin-orbital coupling (SOC), and upon turning on strong SOC, each nodal line is either split or gapped due to the hybridization between opposite spin components\cite{Weng2015,Zeng2015}. It is natural to ask if there is another class of nodal lines with nontrivial monopole charges, and if there are nodal lines that are robust even in the presence of SOC, i.e., four-band crossing lines. These open questions motivate us to develop a more comprehensive theory on TLSM.

Our results are presented in two parts on TLSMs without and with SOC, respectively. In the first part, we revisit systems with $P$ and $T$ in the absence of SOC, and find a new class of nodal lines that can only be created and annihilated in pairs, characterized by a new $Z_2$ topological invariant. For a closed surface around the nodal line, we define a new $Z_2$ invariant protected by $P*T$, classifying all nodal lines into two classes, with and without a $Z_2$-charge, respectively. Nodal lines with $Z_2$ charge can only be created and annihilated in pairs, as the total charge of the BZ must be zero. Finite perturbation can make a nodal line with $Z_2$ charge shrink to an accidental nodal point, but cannot gap it. In the second part, we discuss systems with $P$, $T$ and strong SOC. We show that if there is an additional twofold screw axis, a four-band crossing line, or a double nodal line (crossing between two doubly degenerate bands), can be protected on the boundary of the BZ. This is the first analytic proof of the symmetry protection of a four-band crossing line. We apply the resultant theory to explain the double nodal line found in earlier model studies\cite{Carter2012,Chen2015} on SrIrO$_3$.

The symmetry we consider is the composition of $P$ and $T$, or $P*T$, an anti-unitary symmetry that preserves the momentum of a single particle. In a system without SOC, we have $T^2=1$, $P^2=1$ and $[P,T]=0$, which imply $(P*T)^2=+1$. The action of $P*T$ on the atomic orbitals can hence be represented by complex conjugation ($K$) up to a basis choice. Therefore, the single particle Hamiltonian, $H(\bk)$, is a real matrix at every $\bk$ in BZ. A real, gapped Hamiltonian has $Z_2$ topological classification in both one and two dimensions, indicated by the first and the second homotopy groups of the projector onto the occupied bands\cite{Hatcher2002}
\bea
\pi_1(\frac{O(M+N)}{O(M)\oplus{O}(N)})=\pi_2(\frac{O(M+N)}{O(M)\oplus{O}(N)})=Z_2,
\eea
where $M$ and $N$ are the numbers of the unoccupied and the occupied bands ($M,N>1$).
Since $H(\bk)$ is real, at each $\bk$ there is a real representation for all eigenstates of $H(\bk)$. Therefore for each projector onto the occupied bands, $P(\bk)$, $1-2P(\bk)$ is an $O(M+N)$ matrix that is invariant under any rotation within the occupied (unoccupied) space, i.e., an element of the quotient group.

\begin{figure}[tbp]
\includegraphics[width=0.45\textwidth]{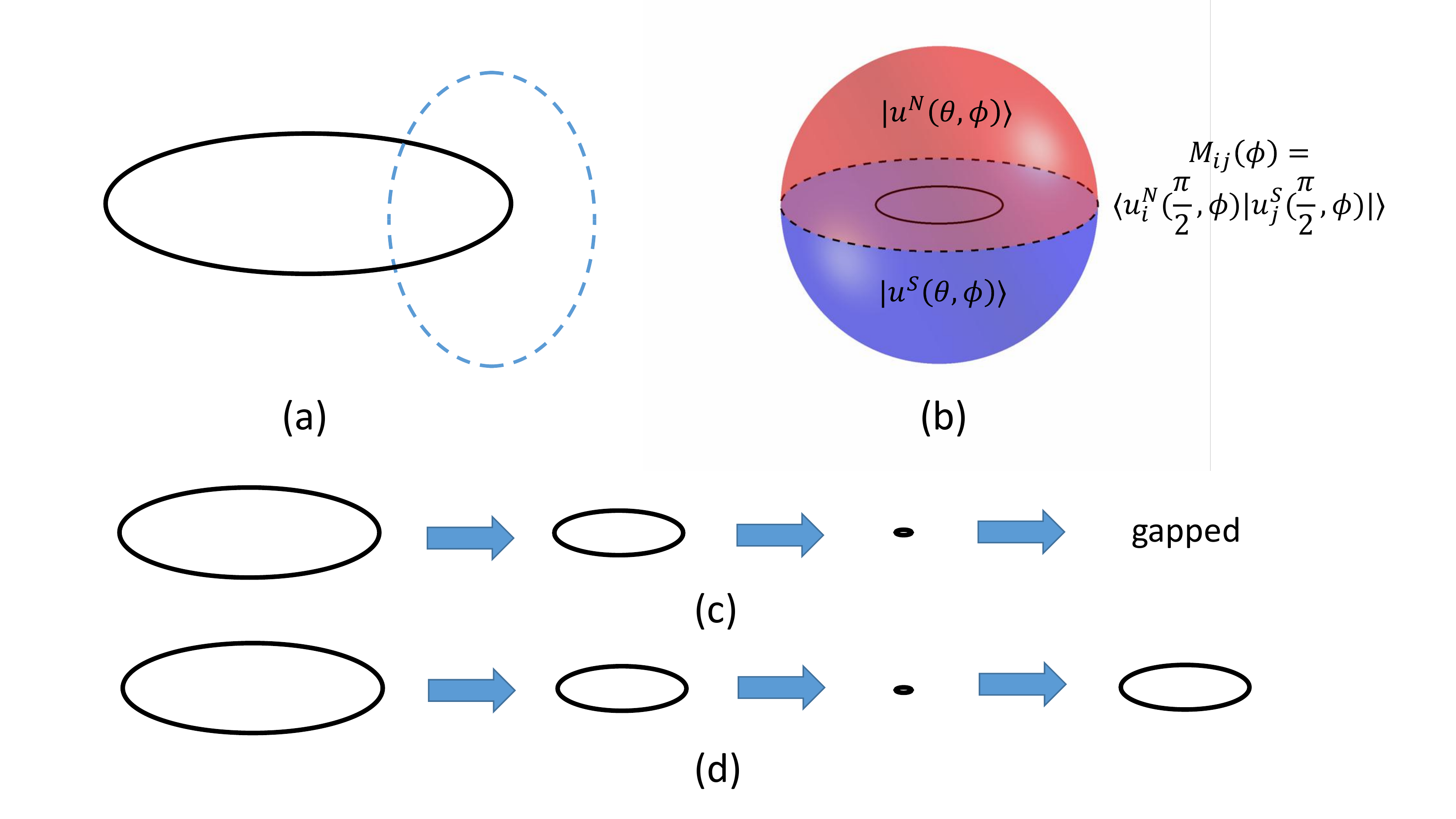}
\caption{(a) A nontrivial $Z_2$ invariant (Berry's phase of $\pi$) of any loop in 3D BZ implies a nodal line (solid line) passing through the loop. (b) The 2D $Z_2$ invariant for a nodal line in 3D BZ defined on a sphere enclosing the line. (c) The evolution of a nodal line with zero monopole charge as parameter changes in the model of Eq.(\ref{eq:model1}). (d) The evolution of a nodal line with nonzero monopole charge as parameter changes in the model of Eq.(\ref{eq:model2})}.
\label{fig:1}
\end{figure}

The $Z_2$ classification of $H(\bk)$ in 1D directly leads to protected nodal points in 2D and nodal lines in 3D, which have been studied in Refs.[\onlinecite{Burkov2011b,Weng2014,Kim2015}]. To see this, consider a closed path, i.e., a loop, in the 3D BZ, along which the Hamiltonian is gapped. The $Z_2$ invariant for the loop is simply the Berry's phase for all occupied bands, quantized to either $0$ or $\pi$, corresponding to the trivial and the nontrivial classes respectively. If a loop belongs to the nontrivial $Z_2$ class, it cannot shrink to a point and vanish without crossing a singularity. In 3D BZ, this implies a line of singularities threading through the loop [see Fig.\ref{fig:1}(a)]. Given $H(\bk)$ continuous, each singularity is where the gap closes between the conduction and the valence bands, i.e., a nodal point, and a line of singularities is hence a nodal line. Given a nodal line in 3D, any loop that interlocks with the nodal line has Berry's phase of $\pi$, while the other loops have zero Berry's phase. Thus we conclude that $P*T$ topologically protects a nodal line in 3D in the absence of SOC. Unlike the monopole charge of Weyl nodes, the 1D $Z_2$ invariant does not forbid a single nodal line from being annihilated or created locally in $k$-space in 3D. Consider the following two band model as example
\bea\label{eq:model1}
H(\bk)=(m-k^2)\sigma_x+k_z\sigma_z.
\eea
$H(\bk)$ has a nodal line on the $k_z=0$ plane of radius $\sqrt{m}$ if $m>0$. As we change $m$ from positive to negative, the nodal line shrinks to a point at the origin and vanishes [see Fig.\ref{fig:1}(c)]. The reverse process creates a single nodal line from the origin. This is a key difference between this nodal line and a point node in Weyl semimetal or some Dirac semimetal. In the latter cases, each point node has a monopole charge, and therefore can only be created or annihilated in pairs.

The nontrivial $Z_2$ classification for real, gapped Hamiltonians in 2D indicates a new classification for nodal lines in 3D. It implies that with $P*T$, $H(\bk)$ on the surface of a sphere can be topologically nontrivial. In 3D $k$-space, if $H(\bk)$ on the surface of a sphere belongs to the nontrivial class, the sphere cannot shrink to a point and vanish without meeting a singularity. Since $P*T$ cannot stabilize a point node, the singularity is a nodal line inside the sphere.

The 2D $Z_2$ invariant for nodal lines in 3D $k$-space, defined on a surface enclosing the line [see Fig.\ref{fig:1}(b)], is a new topological invariant. It indicates the absence/presence of the obstruction to finding a smooth and \emph{real} gauge for the periodic part of the Bloch wavefunctions (wavefunctions for short hereafter) of the occupied bands. This invariant can be constructed based on this observation. Consider a sphere divided by an equator into two halves, say the northern and the southern hemispheres. The wavefunctions on each hemisphere, $|u^{N}_n(\theta,\phi)\rangle$ and $|u^S_n(\theta,\phi)\rangle$, can be real and smooth because each hemisphere is contractible. At each point on the equator, denoted by the azimuthal angle $\phi$, where the two hemispheres meet, the projectors to the occupied space must be equal
\bea\nonumber
\sum_{n\in{occ.}}|u^N_n(\frac{\pi}{2},\phi)\rangle\langle{u}^N_n(\frac{\pi}{2},\phi)|=\sum_{n\in{occ.}}|u^S_n(\frac{\pi}{2},\phi)\rangle\langle{u}^S_n(\frac{\pi}{2},\phi)|,
\eea
because the Hamiltonian is smooth on the whole sphere. Therefore the matrix
\bea
M_{mn}(\phi)=\langle{u}^N_m(\frac{\pi}{2},\phi)|u^S_n(\frac{\pi}{2},\phi)\rangle
\eea
is an $N_{occ}$-by-$N_{occ}$ orthogonal matrix. For the orthogonal group, there is
\bea
\pi_1[O(N_{occ})]=Z_2.
\eea
When $M(\phi)$ is $Z_2$ nontrivial, there is an obstruction to defining a smooth gauge on the whole sphere. In Appendix \ref{apn:explicit}, we show how to obtain this invariant without using a smooth gauge on hemispheres.

The new $Z_2$ invariant classifies nodal lines in 3D systems with $P*T$ into two classes: with and without a $Z_2$ charge. A nodal line with a $Z_2$ charge can be considered a $Z_2$ monopole, which can only be created or annihilated in pairs. This is easy to prove by contradiction: if a $Z_2$ monopole is created (annihilated) locally in $k$-space, the wavefunctions sufficiently away from this point are changed by a small amount, so the $Z_2$ invariant on a surface far away from the point is unchanged, contradicting the assumption that a $Z_2$ monopole is created (annihilated) within.

For concreteness, we construct a 3D Hamiltonian around such a $Z_2$ monopole
\bea\label{eq:model2}
H(\bk)=q_xs_x+q_y\tau_ys_y+q_zs_z+m\tau_xs_x,
\eea
where $\tau_i$ and $s_i$ are Pauli matrices acting on two isospin degrees of freedom and $\bq\equiv\bk-\bk_0$ is the momentum relative to the origin of the $k\cdot{p}$ expansion. The spectrum is given by
\bea
E(\bk)=\pm\sqrt{q_z^2+(\sqrt{q^2_x+q^2_y}\pm{m})^2}.
\eea
The band crossing can be found by solving $E(\bq)=0$, yielding $k_z=0$ and $\sqrt{q_x^2+q_y^2}=|m|$, i.e., a nodal line on the $xy$-plane of radius $\sqrt{|m|}$. As $m$ changes from positive to negative, the radius decreases and shrinks to zero at $m=0$ but increases again when $m$ becomes negative [see Fig.\ref{fig:1}(d)]. An explicit calculation of the $Z_2$ charge of this nodal line is given in Appendix \ref{apn:model2}.

We emphasize that the SU(2) rotation plays an important role in protecting a line node. When SU(2) is broken, or there is SOC, the composition of inversion and time-reversal ensures double degeneracy at each $\bk$ in BZ. Any crossing point between two doublet bands is hence a four-band crossing; yet it is easy to see that a four-band crossing is not protected in any 3D system without additional symmetries. Without SU(2), the symmetry $P*T$ satisfies $(P*T)^2=-1$, because $T^2=-1$ in a spin half system . A generic four-band model with $P*T$ in the presence of SOC is
\bea\label{eq:kdotp}
H(\bk)=f_1s_x+f_2\tau_ys_y+f_3s_z+f_4\tau_xs_y+f_5\tau_zs_y,
\eea
where each $f_i$ is a function of $(k_x,k_y,k_z)$, $P*T=K(i\tau_y)$. A band crossing requires five equations to be satisfied, namely, $f_1=f_2=f_3=f_4=f_5=0$, which is impossible in a 3D BZ without fine tuning. It is natural to ask if additional symmetries can protect a four-band crossing nodal line in 3D systems with SOC, and further if yes, what are they.

Recent work shows that in the presence of nonsymmorphic symmetries, four-band crossings can appear at high symmetry 0D nodal points on the BZ boundary\cite{Young2015}. A nonsymmorphic symmetry is a point group symmetry composed with a fractional lattice translation; the commutation relation involving a nonsymmorphic symmetry is generally $k$-dependent due to the translation part\cite{Parameswaran2013,Liu2014b,Fang2015,Shiozaki2015,Watanabe2015}. At $\Gamma$, a nonsymmorphic symmetry can be treated as its point group component as long as the commutation relations are concerned; at BZ boundary, the fractional translation makes the group structure of nonsymmorphic symmetries different from any point group, and leads to new types of band crossings and high degeneracies\cite{Bradley2010}.

In this paper we show that the presence of $P$, $T$ and a twofold screw axis protect double nodal lines (four-band crossing lines) on the BZ boundary in a 3D system with SOC. Unlike $P$ that only acts in the real space, a twofold screw axis along $z$ ($S_z$) acts in both the real space $(x,y,z)$ and the spin space $(s_x,s_y,s_z)$ simultaneously:
\bea
S_z:&&(x,y,z)\rightarrow(-x+\frac{\mu{a}}{2},-y+\frac{{\lambda}b}{2},z-\frac{c}{2})\\
\nonumber
&&(s_x,s_y,s_z)\rightarrow(-s_x,-s_y,s_z),
\eea
where $\mu,\lambda=0,1$ denoting the shift of the axis, $(\mu{a}/4,\lambda{b}/4,z)$, from the inversion center and $a$, $b$ and $c$ are the lengths of three basis vectors. Combining the twofold axis and inversion generates the another symmetry:
\bea
R_z:&&(x,y,z)\rightarrow(x-\frac{\mu{a}}{2},y-\frac{{\lambda}b}{2},-z+c/2)\\
\nonumber
&&(s_x,s_y,s_z)\rightarrow(-s_x,-s_y,s_z).
\eea
$R$ is a mirror plane (if $\mu=\lambda=0$) or a glide plane (if $\mu$ or $\lambda$ is nonzero) located at $z=c/4$.
Consider the commutation relation between $P*T$ and $R_z$. In real space we have
\bea
(x,y,z,t)&\underrightarrow{R_z}&(x-\frac{\mu{a}}{2},y-\frac{\lambda{b}}{2},-z+c/2,t)\\
\nonumber&\underrightarrow{P*T}&(-x+\frac{\mu{a}}{2},-y+\frac{\lambda{b}}{2},z-c/2,-t),\\
\nonumber
(x,y,z,t)&\underrightarrow{P*T}&(-x,-y,-z,-t)\\
\nonumber
&\underrightarrow{R_z}&(-x-\frac{\mu{a}}{2},-y-\frac{\lambda{b}}{2},z+c/2,-t),
\eea
and in spin space
\bea
(s_x,s_y,s_z)&\underrightarrow{R_z}&(-s_x,-s_y,s_z)\\
\nonumber&\underrightarrow{P*T}&(s_x,s_y,-s_z),\\
\nonumber
(s_x,s_y,s_z)&\underrightarrow{P*T}&(-s_x,-s_y,-s_z)\\
\nonumber
&\underrightarrow{R_z}&(s_x,s_y,-s_z),
\eea
from which we find
\bea\label{eq:commu}
R_z*(P*T)&=&T_{(-\mu{a},-\lambda{b},c)}(P*T)*R_z\\
\nonumber&=&e^{-ik_z+i{\mu}k_x+i{\lambda}k_y}(P*T)*R_z.
\eea
There are two planes (mirror invariant planes) defined by $k_z=0$ and $k_z=\pi$ in the BZ that are invariant under $R_z$, on which the commutation relations given by Eq.(\ref{eq:commu}) differ by a minus sign.

\begin{figure}[tbp]
\includegraphics[width=0.45\textwidth]{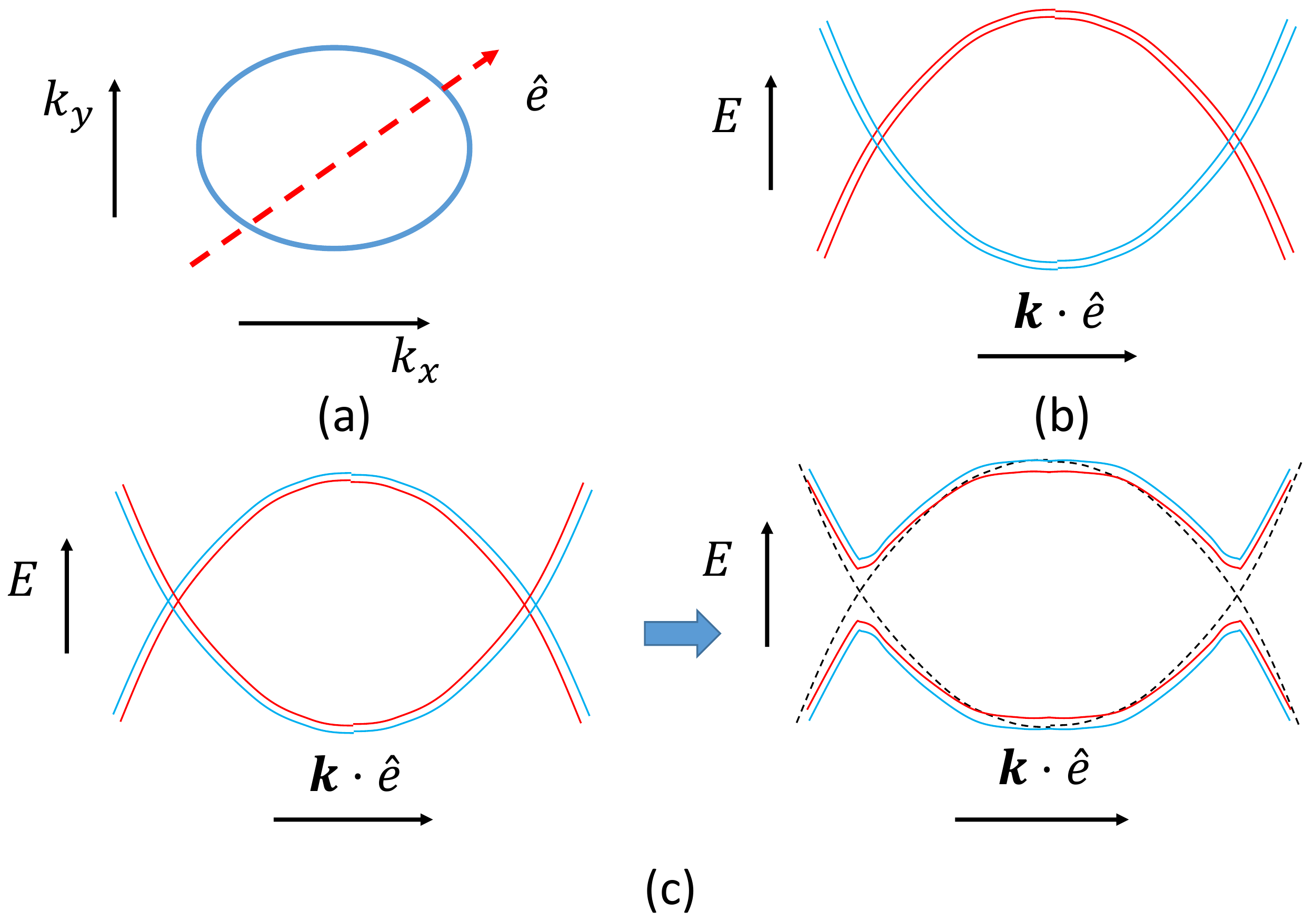}
\caption{(a) A band crossing line on some mirror invariant plane at $k_z=0$ or $k_z=\pi$, with an arbitrary cut along which the band structure is plotted. (b) The band structure along the cut in (a) on $k_z=\pi$, with the corresponding eigenvalues of $R_z$ shown, where the degenerate states of a doublet band have the same eigenvalue. (c) The same band structure on $k_z=0$, where the degenerate states of a doublet band have opposite $R_z$ eigenvalues, which can anti-cross with another doublet band due to the spin mixing enabled by SOC.}
\label{fig:2}
\end{figure}

On each mirror invariant plane, the bands can be labeled by their respective $R_z$ eigenvalues. In a system with SOC, we have
\bea
R_z^2:&&(x,y,z)\rightarrow(x-\mu{a},y-\lambda{b},z),\\
\nonumber
&&(s_x,s_y,s_z)\rightarrow(s_x,s_y,s_z).
\eea
or
\bea
R_z^2=-T_{-\mu{a},-\lambda{b},0}=-e^{i\mu{k_x}+i\lambda{k_y}}.
\eea
The minus sign is because $R_z$ is equivalent to a $\pi$-rotation along $z$ in the spin space so $R^2_z$ includes a $2\pi$-rotation, giving a $-1$ for a spin-$1/2$ system. Therefore, each band at $k_z=0$ and $k_z=\pi$ either has $R_z$ eigenvalue $g_+=+ie^{(i\mu{k_x}+i\lambda{k_y})/2}$ or $g_-=-g_+$. In the presence of SOC and $P*T$, bands are doubly degenerate, and the degenerate bands are related to each other by $P*T$. Suppose at $(k_x,k_y,k_0)$, where $k_0=0,\pi$, a Bloch function $|\psi(\bk)\rangle$ is an eigenstate of $R_z$ with eigenvalue $g_+$, then we consider its degenerate partner $P*T|\psi(\bk)\rangle$ under $R_z$
\bea\nonumber
R_z(P*T)|\psi(\bk)\rangle&=&e^{-ik_0+i\mu{k_x}+i\lambda{k_y}}(P*T)R_z|\psi(\bk)\rangle\\
\nonumber
&=&e^{-ik_0+i\mu{k_x}+i\lambda{k_y}}P*T{g_+}|\psi(\bk)\rangle\\
\nonumber
&=&e^{-ik_0}g_-P*T|\psi(\bk)\rangle.
\eea
At $k_0=0$, the degenerate bands have opposite $R_z$ eigenvalues, and two sets of such doublet bands generally anti-cross: the bands with the same $R_z$ eigenvalue hybridize and avoid crossing [see Fig.\ref{fig:2}(c)]. At $k_0=\pi$, however, the degenerate bands have the same $R_z$ eigenvalue. In this case, two doublet bands with opposite $R_z$ eigenvalues may cross each other along a nodal line, making a symmetry protected four-band crossing line, or a double nodal line [see Fig.\ref{fig:2}(b)]. In Appendix \ref{apn:kdotp}, we revisit the 3D $k.p$-model in Eq.(\ref{eq:kdotp}) in the presence of $R_z$ in addition to $P*T$ and show the presence of a double nodal line; and in Sec.~IV, we write down a formal invariant for the double nodal lines.

We apply the theory for double nodal lines to the case of the iridate SrIrO$_3$. An eight-band tight-binding model consistent with all symmetries in the space group has been shown to exhibit the double line\cite{Carter2012,Rhim2015,Chen2015} around point $U$ in BZ, but a general symmetry analysis for arbitrary number of bands is missing to pin down which symmetries of the little group at $U$ are protecting the double nodal line, whereas the other symmetries may be broken without opening a gap or splitting the double line into two single nodal lines and/or point nodes. The little group at $U$ is generated by $P$ plus two screw axes:
\bea
P: (x,y,z)&\rightarrow&(-x,-y,-z),\\
\nonumber
S_y: (x,y,z)&\rightarrow&(-x+a/2,y+b/2,-z+c/2),\\
\nonumber
S_z: (x,y,z)&\rightarrow&(-x,-y,z+c/2).
\eea
Following our theory, we see that $P$, $T$ and $S_y$ ($S_z$) can protect double nodal lines on the $k_y=\pi$ ($k_z=\pi$) plane. Therefore, the double nodal line predicted in Ref.[\onlinecite{Carter2012}] on $k_y=\pi$ plane is protected by $P$, $T$ and $S_y$. The other symmetries, including $S_z$, $S_x=S_y*S_z$, $M_z=S_z*P$ and $G_b=P*S_x$, may all be broken without gapping or splitting the double line. These statements may be tested in future experiments, as the crystalline symmetries can be modified by applying epitaxial strain on thin-film samples\cite{Liu2015}.

We briefly comment on the possibility of surface states in the TLSMs proposed. The protection of the nodal lines in this work requires the presence of $P$, as opposed to the Weyl semimetals where no symmetry (other than translation) is required. Since an open surface always breaks the inversion, no protected surface states, strictly speaking, are associated with these TLSMs; however, as shown in Ref.[\onlinecite{Chen2015,Weng2014}], when the conduction and the valence bands are nearly symmetric, there is a nearly flat surface band bounded by the projection of the nodal line\footnote{The surface flat band proposed in SrIrO$_3$ is protected by the mirror symmetry $M_z$ and an approximate chiral symmetry specific to this system\cite{Chen2015}. The chiral symmetry makes the conduction and the valence bands symmetric in energy.}.

We propose two new classes of three-dimensional topological semimetals with nodal lines in systems with and without spin-orbital coupling. Without spin-orbital coupling, or equivalently, with SU(2) spin rotation symmetry, inversion and time-reversal can protect a nodal line that carries a $Z_2$ monopole charge, independent of the previously known $\pi$ Berry's phase. Nodal lines with a nonzero monopole charge can only be created or annihilated in pairs, while the nodal lines in previous studies can be singly created and annihilated. In the presence of SOC, we prove that inversion plus time-reversal are insufficient to protect any band crossing in 3D, and an additional nonsymmorphic symmetry (twofold screw axis) can protect a double nodal line, where two sets of doubly degenerate bands cross each other. We apply the theory to SrIrO$_3$, and identify symmetries required to protect the four-band crossing found in earlier model studies, and also symmetries that can be broken without gapping or splitting the double nodal line.

$^\ast$fangc@mit.edu

\acknowledgements{CF thanks A. Alexandradinata and B.~A. Bernevig for useful discussion on the Wilson line technique used in the Supplementary Materials, and CF thanks C.-K. Chiu and S. Ryu for early discussion on the homotopy groups. CF and LF were supported by S3TEC Solid State Solar Thermal Energy Conversion Center, an Energy Frontier Research Center funded by the U.S. Department of Energy (DOE), Office of Science, Basic Energy Sciences (BES), under Award No. DE-SC0001299/DE-FG02-09ER46577. YC and HYK was supported by the NSERC of Canada and the centre for Quantum Materials at the University of Toronto.}

\begin{appendix}

\section{The explicit $Z_2$ invariant for a gapped and real Hamiltonian in 2D without using a smooth gauge}
\label{apn:explicit}
In this Appendix, we explicitly construct the $Z_2$ invariant for a smooth mapping from a two-sphere to the manifold of all occupied states of a gapped and real Hamiltonian, $O(M+N)/O(M)\oplus{O}(N)$, without using smooth gauges on the northern and the southern hemispheres.

In this construction, we only need fix the gauges on the north pole and the south pole, where the wavefunctions of the occupied states are $|u_{NP,i}\rangle$ and $u_{SP,i}\rangle$ respectively, where $i$ labels the occupied bands and both wavefunctions are real. For each azimuthal angle $\phi\in[0,2\pi)$, there is a half longitude connecting the two poles, $L(\phi)$. Consider the following path-ordered integral
\bea
W(\phi)=P\exp[\int_0^\pi{d}{P}(\theta,\phi)\partial_\theta{P}(\theta,\phi)d\theta],
\eea
or, in the discretized form
\bea
W(\phi)=\lim_{N\rightarrow\infty}\prod_{j=0,...,N}P(j\pi/N,\phi),
\eea
where $P(\theta,\phi)$ is the projector onto the occupied states, namely,
\bea
P(\theta,\phi)\equiv\sum_{i=1,...,N_{occ}}|u_i(\theta,\phi)\rangle\langle{u}_i(\theta,\phi)|.
\eea
The physical meaning of $W(\phi)$ is a parallel transport between the occupied space at the north pole and the occupied space at the south pole. Therefore, the following matrix must be unitary
\bea
M_{ij}(\phi)\equiv\langle{u}_{SP,i}|W(\phi)|u_{NP,j}\rangle.
\eea
Due to the reality of $|u_{NP,SP}\rangle$ and of the Hamiltonian, $M_{ij}(\phi)$ is also real and hence orthogonal. $M_{ij}(\phi)$ is hence a smooth mapping from a closed path to the space of orthogonal transforms. Due to $\pi_1[O(N)]=Z_2$, $M_{ij}(\phi)$ can either be trivial or nontrivial, and this $Z_2$ invariant of $M(\phi)$ is the $Z_2$ invariant of $H(\theta,\phi)$.

Given a matrix function $M(\phi)$ in $O(N_{occ})$, how do we know if it is trivial or nontrivial? Consider a unit vector in the real vector space of $N_{occ}$ dimensions: $\mathbf{n}_0=(0,0,...,1)^T$. Acting $M(\phi)$ on this vector results in a path in $S^{N_{occ}-1}$:
\bea
\mathbf{n}(\phi)=M(\phi)\mathbf{n}_0.
\eea
Define
\bea
\mathbf{n}'(\phi)\equiv\frac{\mathbf{n}(\phi)-(\mathbf{n}(\phi)\cdot\mathbf{n}_0)\mathbf{n}_0}{|\mathbf{n}(\phi)-(\mathbf{n}(\phi)\cdot\mathbf{n}_0)\mathbf{n}_0|},\\
\nonumber
\alpha(\phi)=\cos^{-1}[\mathbf{n}_0\cdot\mathbf{n}(\phi)].
\eea
Here $\mathbf{n}'(\phi)$ is the unit vector that is perpendicular to $\mathbf{n}_0$ and inside the plane spanned by $\mathbf{n}_0$ and $\mathbf{n}(\phi)$, and $\alpha(\phi)$ is the angle between $\mathbf{n}(\phi)$ and $\mathbf{n}_0$. Define the rotation
\bea
R(\phi,t)&=&\cos[\alpha(\phi)t]+\sin[\alpha(\phi)t]\mathbf{n}_0\mathbf{{n}'}^T(\phi)\\
\nonumber
&-&\sin[\alpha(\phi)t]\mathbf{{n}'}(\phi)\mathbf{n}^T_0,
\eea
where $t\in[0,1]$.
Under these definitions, the orthogonal matrix
\bea
O(\phi,t)=R(\phi,t)M(\phi),
\eea
is a smooth interpolation between $M(\phi)$ at $t=0$ and $M_0(\phi)$ at $t=1$ that satisfies
\bea\label{eq:block}
M_0(\phi)\mathbf{n}_0=\mathbf{n}_0.
\eea
From Eq.(\ref{eq:block}), we see that $M_0(\phi)$ is an orthogonal matrix that is block diagonal: the entries in last row and in the last column are all zero except the unity on the diagonal. Therefore, the first $N_{occ}-1$ columns and rows of $M_0(\phi)$ is an orthogonal matrix in $N_{occ}-1$ dimensional real vector space. This block is defined as the new $M(\phi)$ and we repeat the process. On each step, we smoothly connect $M(\phi)$ to $M_0(\phi)$ that leaves the last unit vector invariant, then reduce the dimension of the matrix by one. This can proceed as long as $dim(M)>2$, because the acting $M(\phi)$ on $\mathbf{n}_0$ gives a closed path in $S^{dim(M)-1}$, and since $\pi_1(S^{n>1})=0$, this path can always shrink to a point smoothly.

When $M(\phi)$ is reduced to an $O(2)$ matrix, we define the U(1) number
\bea
\beta(\phi)=M_{11}(\phi)+iM_{12}(\phi).
\eea
Due to the periodicity of $M(\phi)$, we can define the winding number of $\beta(\phi)$
\bea
2\pi{n}_w=-i\int\beta^\ast(\phi)\partial_\phi\beta(\phi)d\phi.
\eea
The $Z_2$ invariant is the parity of $n_W$, or,
\bea
\xi=\textrm{mod}(n_w,2).
\eea

\section{The $Z_2$ invariant for Eq.(5)}
\label{apn:model2}
The $Z_2$ invariant for any surface enclosing the origin can also be calculated for the model in Eq.(5). We will set $m=0$ to calculate the invariant, and since it is a quantized, a small nonzero $m$ does not change the result. Given a sphere around the origin, the smooth wavefunctions of the occupied bands on the northern and the southern hemispheres are
\bea
|u^N_1\rangle&=&(\cos\phi\sin\frac{\theta}{2},-\cos\frac{\theta}{2},\sin\frac{\theta}{2}\sin\phi,0)^T,\\
\nonumber
|u^N_2\rangle&=&(\sin\frac{\theta}{2}\sin\phi,0,-\sin\frac{\theta}{2}\cos\phi,\cos\frac{\theta}{2})^T,\\
\nonumber
|u^S_1\rangle&=&(\sin\frac{\theta}{2},-\cos\frac{\theta}{2}\cos\phi,0,\cos\frac{\theta}{2}\sin\phi)^T,\\
\nonumber
|u^S_2\rangle&=&(0,\cos\frac{\theta}{2}\sin\phi,-\sin\frac{\theta}{2},\cos\frac{\theta}{2}\cos\phi)^T.
\eea
The connection matrix on the equator is found to be
\bea
M(\phi)=-\cos\phi-i\sin\phi\sigma_2.
\eea
This path corresponds to the nontrivial element of $\pi_1[O(2)]$ as the winding number is $+1$, hence the $Z_2$ invariant is nontrivial.

\section{Double nodal line in a four-band $k\cdot{p}$ model with $P$, $T$ and a twofold screw axis}
\label{apn:kdotp}
We revisit the 3D $k.p$-model in Eq.(7), in the presence of $R_z$ in addition to $P*T$.  When the origin of the $k.p$ expansion is at $\mathbf{k}=(0,0,0)$ ($\mathbf{k}=(0,0,\pi$), we have $[R_z,P*T]=0$ ($\{R_z,P*T\}=0$). On At $\mathbf{k}=(0,0,0)$, the $R_z$ is presented by $R_z=is_y$ up to a gauge. $R_z$ puts constraints on the forms of $f_{i=1,2,3,4}$:
\bea\label{eq:constraints1}
f_{1,3}(q_x,q_y,q_z)&=&-f_{1,3}(q_x,q_y,-q_z),\\
\nonumber
f_{2,4,5}(q_x,q_y,q_z)&=&f_{2,4,5}(q_x,q_y,q_z).
\eea
Eqs.(\ref{eq:constraints1}) guarantee that $f_{1,3}$ vanish on $q_z=0$ plane. But it takes fine tuning to make $f_{2,4,5}$ vanish, i.e., to obtain a band crossing. This is consistent with our previous result that when $[R_z,P*T]=0$, there is no symmetry protected band crossing. If $\{R,P*T\}=0$, or at $\mathbf{k}=(0,0,\pi)$, $R_z$ is represented by $R_z=is_x$ up to a gauge and the constraints on $f_{i=1,2,3,4,5}$ by $R_z$ are
\bea\label{eq:constraints2}
f_1(q_x,q_y,q_z)&=&f_1(q_x,q_y,-q_z),\\
\nonumber
f_{2,3,4,5}(q_x,q_y,q_z)&=&-f_{2,3,4,5}(q_x,q_y,-q_z).
\eea
Eqs.(\ref{eq:constraints2}) guarantee that $f_{2,3,4,5}$ vanish on $q_z=0$ plane. Therefore, in order to obtain a band crossing on this plane, we only need $f_1(q_x,q_y,0)=0$, the solution space to which is generically a 1D line, i.e., a double nodal line.

\section{Topological invariant for a double nodal line}
\label{apn:inv}
According to the text, the double line crossing is formed by two doublet bands that have opposite eigenvalues of $R\equiv{P}*S_z$ (where $S_z$ can be replaced by any twofold screw axis). For a given double nodal ring on the $k_z=\pi$-plane, we can find two points inside and outside the ring on the same plane, denoted by A and B. At A (B), each band is doubly degenerate, where the two degenerate states have the same eigenvalue of $R$, labeled by $\xi_n(A)$ ($\xi_n(B)$). We can hence define the following invariant
\bea
\zeta=\prod_{2n\in{occ}}\frac{\xi_{2n}(A)}{\xi_{2n}(B)}.
\eea
Since $R$ only has eigenvalues of $g_+$ and $g_-=-g_+$, $\zeta=\pm1$ is a $Z_2$ invariant. When $\zeta=1$, the double nodal line between A and B is trivial and can be gapped, and if $\zeta=-1$, the line is protected.
\end{appendix}


\begin{thebibliography}{48}
\expandafter\ifx\csname natexlab\endcsname\relax\def\natexlab#1{#1}\fi
\expandafter\ifx\csname bibnamefont\endcsname\relax
  \def\bibnamefont#1{#1}\fi
\expandafter\ifx\csname bibfnamefont\endcsname\relax
  \def\bibfnamefont#1{#1}\fi
\expandafter\ifx\csname citenamefont\endcsname\relax
  \def\citenamefont#1{#1}\fi
\expandafter\ifx\csname url\endcsname\relax
  \def\url#1{\texttt{#1}}\fi
\expandafter\ifx\csname urlprefix\endcsname\relax\def\urlprefix{URL }\fi
\providecommand{\bibinfo}[2]{#2}
\providecommand{\eprint}[2][]{\url{#2}}

\bibitem[{\citenamefont{Wan et~al.}({2011})\citenamefont{Wan, Turner,
  Vishwanath, and Savrasov}}]{Wan2011}
\bibinfo{author}{\bibfnamefont{X.}~\bibnamefont{Wan}},
  \bibinfo{author}{\bibfnamefont{A.~M.} \bibnamefont{Turner}},
  \bibinfo{author}{\bibfnamefont{A.}~\bibnamefont{Vishwanath}},
  \bibnamefont{and} \bibinfo{author}{\bibfnamefont{S.~Y.}
  \bibnamefont{Savrasov}}, \bibinfo{journal}{Phys. Rev. B}
  \textbf{\bibinfo{volume}{{83}}}, \bibinfo{pages}{205101}
  (\bibinfo{year}{{2011}}).

\bibitem[{\citenamefont{Hosur et~al.}(2012)\citenamefont{Hosur, Parameswaran,
  and Vishwanath}}]{Hosur2012}
\bibinfo{author}{\bibfnamefont{P.}~\bibnamefont{Hosur}},
  \bibinfo{author}{\bibfnamefont{S.~A.} \bibnamefont{Parameswaran}},
  \bibnamefont{and}
  \bibinfo{author}{\bibfnamefont{A.}~\bibnamefont{Vishwanath}},
  \bibinfo{journal}{Phys. Rev. Lett.} \textbf{\bibinfo{volume}{108}},
  \bibinfo{pages}{046602} (\bibinfo{year}{2012}).

\bibitem[{\citenamefont{Son and Spivak}(2013)}]{Son2013}
\bibinfo{author}{\bibfnamefont{D.~T.} \bibnamefont{Son}} \bibnamefont{and}
  \bibinfo{author}{\bibfnamefont{B.~Z.} \bibnamefont{Spivak}},
  \bibinfo{journal}{Phys. Rev. B} \textbf{\bibinfo{volume}{88}},
  \bibinfo{pages}{104412} (\bibinfo{year}{2013}).

\bibitem[{\citenamefont{Young et~al.}(2012)\citenamefont{Young, Zaheer, Teo,
  Kane, Mele, and Rappe}}]{Young2012}
\bibinfo{author}{\bibfnamefont{S.~M.} \bibnamefont{Young}},
  \bibinfo{author}{\bibfnamefont{S.}~\bibnamefont{Zaheer}},
  \bibinfo{author}{\bibfnamefont{J.~C.~Y.} \bibnamefont{Teo}},
  \bibinfo{author}{\bibfnamefont{C.~L.} \bibnamefont{Kane}},
  \bibinfo{author}{\bibfnamefont{E.~J.} \bibnamefont{Mele}}, \bibnamefont{and}
  \bibinfo{author}{\bibfnamefont{A.~M.} \bibnamefont{Rappe}},
  \bibinfo{journal}{Phys. Rev. Lett.} \textbf{\bibinfo{volume}{108}},
  \bibinfo{pages}{140405} (\bibinfo{year}{2012}).

\bibitem[{\citenamefont{Burkov et~al.}({2011})\citenamefont{Burkov, Hook, and
  Balents}}]{Burkov2011}
\bibinfo{author}{\bibfnamefont{A.~A.} \bibnamefont{Burkov}},
  \bibinfo{author}{\bibfnamefont{M.~D.} \bibnamefont{Hook}}, \bibnamefont{and}
  \bibinfo{author}{\bibfnamefont{L.}~\bibnamefont{Balents}},
  \bibinfo{journal}{Phys. Rev. B} \textbf{\bibinfo{volume}{{84}}},
  \bibinfo{pages}{235126} (\bibinfo{year}{{2011}}).

\bibitem[{\citenamefont{Burkov and Balents}(2011)}]{Burkov2011a}
\bibinfo{author}{\bibfnamefont{A.~A.} \bibnamefont{Burkov}} \bibnamefont{and}
  \bibinfo{author}{\bibfnamefont{L.}~\bibnamefont{Balents}},
  \bibinfo{journal}{Phys. Rev. Lett.} \textbf{\bibinfo{volume}{107}},
  \bibinfo{pages}{127205} (\bibinfo{year}{2011}).

\bibitem[{\citenamefont{Fang et~al.}(2012)\citenamefont{Fang, Gilbert, Dai, and
  Bernevig}}]{Fang2012}
\bibinfo{author}{\bibfnamefont{C.}~\bibnamefont{Fang}},
  \bibinfo{author}{\bibfnamefont{M.~J.} \bibnamefont{Gilbert}},
  \bibinfo{author}{\bibfnamefont{X.}~\bibnamefont{Dai}}, \bibnamefont{and}
  \bibinfo{author}{\bibfnamefont{B.~A.} \bibnamefont{Bernevig}},
  \bibinfo{journal}{Phys. Rev. Lett.} \textbf{\bibinfo{volume}{108}},
  \bibinfo{pages}{266802} (\bibinfo{year}{2012}).

\bibitem[{\citenamefont{Liu et~al.}(2013)\citenamefont{Liu, Ye, and
  Qi}}]{Liu2013}
\bibinfo{author}{\bibfnamefont{C.-X.} \bibnamefont{Liu}},
  \bibinfo{author}{\bibfnamefont{P.}~\bibnamefont{Ye}}, \bibnamefont{and}
  \bibinfo{author}{\bibfnamefont{X.-L.} \bibnamefont{Qi}},
  \bibinfo{journal}{Phys. Rev. B} \textbf{\bibinfo{volume}{87}},
  \bibinfo{pages}{235306} (\bibinfo{year}{2013}).

\bibitem[{\citenamefont{Lu et~al.}(2013)\citenamefont{Lu, Fu, Joannopoulos, and
  Soljacic}}]{Lu2013}
\bibinfo{author}{\bibfnamefont{L.}~\bibnamefont{Lu}},
  \bibinfo{author}{\bibfnamefont{L.}~\bibnamefont{Fu}},
  \bibinfo{author}{\bibfnamefont{J.~D.} \bibnamefont{Joannopoulos}},
  \bibnamefont{and} \bibinfo{author}{\bibfnamefont{M.}~\bibnamefont{Soljacic}},
  \bibinfo{journal}{Nature Photonics}  (\bibinfo{year}{2013}).

\bibitem[{\citenamefont{Weng et~al.}(2015)\citenamefont{Weng, Fang, Fang,
  Bernevig, and Dai}}]{Weng2015}
\bibinfo{author}{\bibfnamefont{H.}~\bibnamefont{Weng}},
  \bibinfo{author}{\bibfnamefont{C.}~\bibnamefont{Fang}},
  \bibinfo{author}{\bibfnamefont{Z.}~\bibnamefont{Fang}},
  \bibinfo{author}{\bibfnamefont{B.~A.} \bibnamefont{Bernevig}},
  \bibnamefont{and} \bibinfo{author}{\bibfnamefont{X.}~\bibnamefont{Dai}},
  \bibinfo{journal}{Phys. Rev. X} \textbf{\bibinfo{volume}{5}},
  \bibinfo{pages}{011029} (\bibinfo{year}{2015}).

\bibitem[{\citenamefont{Huang et~al.}(2015{\natexlab{a}})\citenamefont{Huang,
  Xu, Belopolski, Lee, Chang, Wang, Alidoust, Bian, Neupane, Bansil
  et~al.}}]{Huang2015}
\bibinfo{author}{\bibfnamefont{S.-M.} \bibnamefont{Huang}},
  \bibinfo{author}{\bibfnamefont{S.-Y.} \bibnamefont{Xu}},
  \bibinfo{author}{\bibfnamefont{I.}~\bibnamefont{Belopolski}},
  \bibinfo{author}{\bibfnamefont{C.-C.} \bibnamefont{Lee}},
  \bibinfo{author}{\bibfnamefont{G.}~\bibnamefont{Chang}},
  \bibinfo{author}{\bibfnamefont{B.}~\bibnamefont{Wang}},
  \bibinfo{author}{\bibfnamefont{N.}~\bibnamefont{Alidoust}},
  \bibinfo{author}{\bibfnamefont{G.}~\bibnamefont{Bian}},
  \bibinfo{author}{\bibfnamefont{M.}~\bibnamefont{Neupane}},
  \bibinfo{author}{\bibfnamefont{A.}~\bibnamefont{Bansil}},
  \bibnamefont{et~al.}, \bibinfo{journal}{arXiv:1501.00755}
  (\bibinfo{year}{2015}{\natexlab{a}}).

\bibitem[{\citenamefont{Wang et~al.}({2012})\citenamefont{Wang, Sun, Chen,
  Franchini, Xu, Weng, Dai, and Fang}}]{Wang2012}
\bibinfo{author}{\bibfnamefont{Z.}~\bibnamefont{Wang}},
  \bibinfo{author}{\bibfnamefont{Y.}~\bibnamefont{Sun}},
  \bibinfo{author}{\bibfnamefont{X.-Q.} \bibnamefont{Chen}},
  \bibinfo{author}{\bibfnamefont{C.}~\bibnamefont{Franchini}},
  \bibinfo{author}{\bibfnamefont{G.}~\bibnamefont{Xu}},
  \bibinfo{author}{\bibfnamefont{H.}~\bibnamefont{Weng}},
  \bibinfo{author}{\bibfnamefont{X.}~\bibnamefont{Dai}}, \bibnamefont{and}
  \bibinfo{author}{\bibfnamefont{Z.}~\bibnamefont{Fang}},
  \bibinfo{journal}{Phys. Rev. B} \textbf{\bibinfo{volume}{{85}}},
  \bibinfo{pages}{195320} (\bibinfo{year}{{2012}}), ISSN
  \bibinfo{issn}{{1098-0121}}.

\bibitem[{\citenamefont{Zeng et~al.}(2015)\citenamefont{Zeng, Fang, Chang,
  Chen, Hsieh, Bansil, Lin, and Fu}}]{Zeng2015}
\bibinfo{author}{\bibfnamefont{M.}~\bibnamefont{Zeng}},
  \bibinfo{author}{\bibfnamefont{C.}~\bibnamefont{Fang}},
  \bibinfo{author}{\bibfnamefont{G.}~\bibnamefont{Chang}},
  \bibinfo{author}{\bibfnamefont{Y.-A.} \bibnamefont{Chen}},
  \bibinfo{author}{\bibfnamefont{T.}~\bibnamefont{Hsieh}},
  \bibinfo{author}{\bibfnamefont{A.}~\bibnamefont{Bansil}},
  \bibinfo{author}{\bibfnamefont{H.}~\bibnamefont{Lin}}, \bibnamefont{and}
  \bibinfo{author}{\bibfnamefont{L.}~\bibnamefont{Fu}},
  \bibinfo{journal}{arXiv:1504.03492}  (\bibinfo{year}{2015}).

\bibitem[{\citenamefont{Lu et~al.}(2015)\citenamefont{Lu, Wang, Ye, Ran, Fu,
  Joannopoulos, and Soljacic}}]{Lu2015}
\bibinfo{author}{\bibfnamefont{L.}~\bibnamefont{Lu}},
  \bibinfo{author}{\bibfnamefont{Z.}~\bibnamefont{Wang}},
  \bibinfo{author}{\bibfnamefont{D.}~\bibnamefont{Ye}},
  \bibinfo{author}{\bibfnamefont{L.}~\bibnamefont{Ran}},
  \bibinfo{author}{\bibfnamefont{L.}~\bibnamefont{Fu}},
  \bibinfo{author}{\bibfnamefont{J.~D.} \bibnamefont{Joannopoulos}},
  \bibnamefont{and} \bibinfo{author}{\bibfnamefont{M.}~\bibnamefont{Soljacic}},
  \bibinfo{journal}{arXiv:1502.03438}  (\bibinfo{year}{2015}).

\bibitem[{\citenamefont{Xu et~al.}(2015{\natexlab{a}})\citenamefont{Xu,
  Belopolski, Alidoust, Neupane, Zhang, Sankar, Huang, Lee, Chang, Wang
  et~al.}}]{Xu2015a}
\bibinfo{author}{\bibfnamefont{S.-Y.} \bibnamefont{Xu}},
  \bibinfo{author}{\bibfnamefont{I.}~\bibnamefont{Belopolski}},
  \bibinfo{author}{\bibfnamefont{N.}~\bibnamefont{Alidoust}},
  \bibinfo{author}{\bibfnamefont{M.}~\bibnamefont{Neupane}},
  \bibinfo{author}{\bibfnamefont{C.}~\bibnamefont{Zhang}},
  \bibinfo{author}{\bibfnamefont{R.}~\bibnamefont{Sankar}},
  \bibinfo{author}{\bibfnamefont{S.-M.} \bibnamefont{Huang}},
  \bibinfo{author}{\bibfnamefont{C.-C.} \bibnamefont{Lee}},
  \bibinfo{author}{\bibfnamefont{G.}~\bibnamefont{Chang}},
  \bibinfo{author}{\bibfnamefont{B.}~\bibnamefont{Wang}}, \bibnamefont{et~al.},
  \bibinfo{journal}{arXiv:1502.03807}  (\bibinfo{year}{2015}{\natexlab{a}}).

\bibitem[{\citenamefont{Lv et~al.}(2015)\citenamefont{Lv, Weng, Fu, Wang, Miao,
  Ma, Richard, Huang, Zhao, Chen et~al.}}]{Lv2015}
\bibinfo{author}{\bibfnamefont{B.~Q.} \bibnamefont{Lv}},
  \bibinfo{author}{\bibfnamefont{H.~M.} \bibnamefont{Weng}},
  \bibinfo{author}{\bibfnamefont{B.~B.} \bibnamefont{Fu}},
  \bibinfo{author}{\bibfnamefont{X.~P.} \bibnamefont{Wang}},
  \bibinfo{author}{\bibfnamefont{H.}~\bibnamefont{Miao}},
  \bibinfo{author}{\bibfnamefont{J.}~\bibnamefont{Ma}},
  \bibinfo{author}{\bibfnamefont{P.}~\bibnamefont{Richard}},
  \bibinfo{author}{\bibfnamefont{X.~C.} \bibnamefont{Huang}},
  \bibinfo{author}{\bibfnamefont{L.~X.} \bibnamefont{Zhao}},
  \bibinfo{author}{\bibfnamefont{G.~F.} \bibnamefont{Chen}},
  \bibnamefont{et~al.}, \bibinfo{journal}{arXiv:1502.04684}
  (\bibinfo{year}{2015}).

\bibitem[{\citenamefont{Zhang et~al.}(2015)\citenamefont{Zhang, Yuan, Xu, Lin,
  Tong, Hasan, Wang, Zhang, and Jia}}]{Zhang2015}
\bibinfo{author}{\bibfnamefont{C.}~\bibnamefont{Zhang}},
  \bibinfo{author}{\bibfnamefont{Z.}~\bibnamefont{Yuan}},
  \bibinfo{author}{\bibfnamefont{S.-Y.} \bibnamefont{Xu}},
  \bibinfo{author}{\bibfnamefont{Z.}~\bibnamefont{Lin}},
  \bibinfo{author}{\bibfnamefont{B.}~\bibnamefont{Tong}},
  \bibinfo{author}{\bibfnamefont{M.~Z.} \bibnamefont{Hasan}},
  \bibinfo{author}{\bibfnamefont{J.}~\bibnamefont{Wang}},
  \bibinfo{author}{\bibfnamefont{C.}~\bibnamefont{Zhang}}, \bibnamefont{and}
  \bibinfo{author}{\bibfnamefont{S.}~\bibnamefont{Jia}},
  \bibinfo{journal}{arXiv:1502.00251}  (\bibinfo{year}{2015}).

\bibitem[{\citenamefont{Huang et~al.}(2015{\natexlab{b}})\citenamefont{Huang,
  Zhao, Long, Wang, Chen, Yang, Liang, Xue, Weng, Fang et~al.}}]{Huang2015a}
\bibinfo{author}{\bibfnamefont{X.}~\bibnamefont{Huang}},
  \bibinfo{author}{\bibfnamefont{L.}~\bibnamefont{Zhao}},
  \bibinfo{author}{\bibfnamefont{Y.}~\bibnamefont{Long}},
  \bibinfo{author}{\bibfnamefont{P.}~\bibnamefont{Wang}},
  \bibinfo{author}{\bibfnamefont{D.}~\bibnamefont{Chen}},
  \bibinfo{author}{\bibfnamefont{Z.}~\bibnamefont{Yang}},
  \bibinfo{author}{\bibfnamefont{H.}~\bibnamefont{Liang}},
  \bibinfo{author}{\bibfnamefont{M.}~\bibnamefont{Xue}},
  \bibinfo{author}{\bibfnamefont{H.}~\bibnamefont{Weng}},
  \bibinfo{author}{\bibfnamefont{Z.}~\bibnamefont{Fang}}, \bibnamefont{et~al.},
  \bibinfo{journal}{arXiv:1503.01304}  (\bibinfo{year}{2015}{\natexlab{b}}).

\bibitem[{\citenamefont{Liu et~al.}({2014})\citenamefont{Liu, Jiang, Zhou,
  Wang, Zhang, Weng, Prabhakaran, Mo, Peng, Dudin et~al.}}]{Liu2014}
\bibinfo{author}{\bibfnamefont{Z.~K.} \bibnamefont{Liu}},
  \bibinfo{author}{\bibfnamefont{J.}~\bibnamefont{Jiang}},
  \bibinfo{author}{\bibfnamefont{B.}~\bibnamefont{Zhou}},
  \bibinfo{author}{\bibfnamefont{Z.~J.} \bibnamefont{Wang}},
  \bibinfo{author}{\bibfnamefont{Y.}~\bibnamefont{Zhang}},
  \bibinfo{author}{\bibfnamefont{H.~M.} \bibnamefont{Weng}},
  \bibinfo{author}{\bibfnamefont{D.}~\bibnamefont{Prabhakaran}},
  \bibinfo{author}{\bibfnamefont{S.-K.} \bibnamefont{Mo}},
  \bibinfo{author}{\bibfnamefont{H.}~\bibnamefont{Peng}},
  \bibinfo{author}{\bibfnamefont{P.}~\bibnamefont{Dudin}},
  \bibnamefont{et~al.}, \bibinfo{journal}{{Nat. Mater}}
  \textbf{\bibinfo{volume}{{13}}}, \bibinfo{pages}{{677}}
  (\bibinfo{year}{{2014}}).

\bibitem[{\citenamefont{Liu et~al.}(2014{\natexlab{a}})\citenamefont{Liu, Zhou,
  Zhang, Wang, Weng, Prabhakaran, Mo, Shen, Fang, Dai et~al.}}]{Liu2014a}
\bibinfo{author}{\bibfnamefont{Z.~K.} \bibnamefont{Liu}},
  \bibinfo{author}{\bibfnamefont{B.}~\bibnamefont{Zhou}},
  \bibinfo{author}{\bibfnamefont{Y.}~\bibnamefont{Zhang}},
  \bibinfo{author}{\bibfnamefont{Z.~J.} \bibnamefont{Wang}},
  \bibinfo{author}{\bibfnamefont{H.~M.} \bibnamefont{Weng}},
  \bibinfo{author}{\bibfnamefont{D.}~\bibnamefont{Prabhakaran}},
  \bibinfo{author}{\bibfnamefont{S.-K.} \bibnamefont{Mo}},
  \bibinfo{author}{\bibfnamefont{Z.~X.} \bibnamefont{Shen}},
  \bibinfo{author}{\bibfnamefont{Z.}~\bibnamefont{Fang}},
  \bibinfo{author}{\bibfnamefont{X.}~\bibnamefont{Dai}}, \bibnamefont{et~al.},
  \bibinfo{journal}{Science}  (\bibinfo{year}{2014}{\natexlab{a}}).

\bibitem[{\citenamefont{Neupane et~al.}(2014)\citenamefont{Neupane, Xu, Sankar,
  Alidoust, Bian, Liu, Belopolski, Chang, Jeng, Lin et~al.}}]{Neupane2014}
\bibinfo{author}{\bibfnamefont{M.}~\bibnamefont{Neupane}},
  \bibinfo{author}{\bibfnamefont{S.-Y.} \bibnamefont{Xu}},
  \bibinfo{author}{\bibfnamefont{R.}~\bibnamefont{Sankar}},
  \bibinfo{author}{\bibfnamefont{N.}~\bibnamefont{Alidoust}},
  \bibinfo{author}{\bibfnamefont{G.}~\bibnamefont{Bian}},
  \bibinfo{author}{\bibfnamefont{C.}~\bibnamefont{Liu}},
  \bibinfo{author}{\bibfnamefont{I.}~\bibnamefont{Belopolski}},
  \bibinfo{author}{\bibfnamefont{T.-R.} \bibnamefont{Chang}},
  \bibinfo{author}{\bibfnamefont{H.-T.} \bibnamefont{Jeng}},
  \bibinfo{author}{\bibfnamefont{H.}~\bibnamefont{Lin}}, \bibnamefont{et~al.},
  \bibinfo{journal}{Nat. Comm.} \textbf{\bibinfo{volume}{5}}
  (\bibinfo{year}{2014}).

\bibitem[{\citenamefont{He et~al.}(2014)\citenamefont{He, Hong, Dong, Pan,
  Zhang, Zhang, and Li}}]{He2014}
\bibinfo{author}{\bibfnamefont{L.~P.} \bibnamefont{He}},
  \bibinfo{author}{\bibfnamefont{X.~C.} \bibnamefont{Hong}},
  \bibinfo{author}{\bibfnamefont{J.~K.} \bibnamefont{Dong}},
  \bibinfo{author}{\bibfnamefont{J.}~\bibnamefont{Pan}},
  \bibinfo{author}{\bibfnamefont{Z.}~\bibnamefont{Zhang}},
  \bibinfo{author}{\bibfnamefont{J.}~\bibnamefont{Zhang}}, \bibnamefont{and}
  \bibinfo{author}{\bibfnamefont{S.~Y.} \bibnamefont{Li}},
  \bibinfo{journal}{Phys. Rev. Lett.} \textbf{\bibinfo{volume}{113}},
  \bibinfo{pages}{246402} (\bibinfo{year}{2014}).

\bibitem[{\citenamefont{Jeon et~al.}(2014)\citenamefont{Jeon, Zhou, Gyenis,
  Feldman, Kimchi, Potter, Gibson, Cava, Vishwanath, and Yazdani}}]{Jeon2014}
\bibinfo{author}{\bibfnamefont{S.}~\bibnamefont{Jeon}},
  \bibinfo{author}{\bibfnamefont{B.~B.} \bibnamefont{Zhou}},
  \bibinfo{author}{\bibfnamefont{A.}~\bibnamefont{Gyenis}},
  \bibinfo{author}{\bibfnamefont{B.~E.} \bibnamefont{Feldman}},
  \bibinfo{author}{\bibfnamefont{I.}~\bibnamefont{Kimchi}},
  \bibinfo{author}{\bibfnamefont{A.~C.} \bibnamefont{Potter}},
  \bibinfo{author}{\bibfnamefont{Q.~D.} \bibnamefont{Gibson}},
  \bibinfo{author}{\bibfnamefont{R.~J.} \bibnamefont{Cava}},
  \bibinfo{author}{\bibfnamefont{A.}~\bibnamefont{Vishwanath}},
  \bibnamefont{and} \bibinfo{author}{\bibfnamefont{A.}~\bibnamefont{Yazdani}},
  \bibinfo{journal}{Nature Materials} \textbf{\bibinfo{volume}{13}},
  \bibinfo{pages}{851} (\bibinfo{year}{2014}).

\bibitem[{\citenamefont{Xu et~al.}(2015{\natexlab{b}})\citenamefont{Xu, Liu,
  Kushwaha, Sankar, Krizan, Belopolski, Neupane, Bian, Alidoust, Chang
  et~al.}}]{Xu2015}
\bibinfo{author}{\bibfnamefont{S.-Y.} \bibnamefont{Xu}},
  \bibinfo{author}{\bibfnamefont{C.}~\bibnamefont{Liu}},
  \bibinfo{author}{\bibfnamefont{S.~K.} \bibnamefont{Kushwaha}},
  \bibinfo{author}{\bibfnamefont{R.}~\bibnamefont{Sankar}},
  \bibinfo{author}{\bibfnamefont{J.~W.} \bibnamefont{Krizan}},
  \bibinfo{author}{\bibfnamefont{I.}~\bibnamefont{Belopolski}},
  \bibinfo{author}{\bibfnamefont{M.}~\bibnamefont{Neupane}},
  \bibinfo{author}{\bibfnamefont{G.}~\bibnamefont{Bian}},
  \bibinfo{author}{\bibfnamefont{N.}~\bibnamefont{Alidoust}},
  \bibinfo{author}{\bibfnamefont{T.-R.} \bibnamefont{Chang}},
  \bibnamefont{et~al.}, \bibinfo{journal}{Science}
  \textbf{\bibinfo{volume}{347}}, \bibinfo{pages}{294}
  (\bibinfo{year}{2015}{\natexlab{b}}).

\bibitem[{\citenamefont{Xiong et~al.}(2015)\citenamefont{Xiong, Kushwaha,
  Krizan, Liang, Cava, and Ong}}]{Xiong2015}
\bibinfo{author}{\bibfnamefont{J.}~\bibnamefont{Xiong}},
  \bibinfo{author}{\bibfnamefont{S.}~\bibnamefont{Kushwaha}},
  \bibinfo{author}{\bibfnamefont{J.}~\bibnamefont{Krizan}},
  \bibinfo{author}{\bibfnamefont{T.}~\bibnamefont{Liang}},
  \bibinfo{author}{\bibfnamefont{R.~J.} \bibnamefont{Cava}}, \bibnamefont{and}
  \bibinfo{author}{\bibfnamefont{N.~P.} \bibnamefont{Ong}},
  \bibinfo{journal}{arXiv:1502.06266}  (\bibinfo{year}{2015}).

\bibitem[{\citenamefont{Burkov et~al.}(2011)\citenamefont{Burkov, Hook, and
  Balents}}]{Burkov2011b}
\bibinfo{author}{\bibfnamefont{A.~A.} \bibnamefont{Burkov}},
  \bibinfo{author}{\bibfnamefont{M.~D.} \bibnamefont{Hook}}, \bibnamefont{and}
  \bibinfo{author}{\bibfnamefont{L.}~\bibnamefont{Balents}},
  \bibinfo{journal}{Phys. Rev. B} \textbf{\bibinfo{volume}{84}},
  \bibinfo{pages}{235126} (\bibinfo{year}{2011}).

\bibitem[{\citenamefont{Carter et~al.}(2012)\citenamefont{Carter, Shankar, Zeb,
  and Kee}}]{Carter2012}
\bibinfo{author}{\bibfnamefont{J.-M.} \bibnamefont{Carter}},
  \bibinfo{author}{\bibfnamefont{V.~V.} \bibnamefont{Shankar}},
  \bibinfo{author}{\bibfnamefont{M.~A.} \bibnamefont{Zeb}}, \bibnamefont{and}
  \bibinfo{author}{\bibfnamefont{H.-Y.} \bibnamefont{Kee}},
  \bibinfo{journal}{Phys. Rev. B} \textbf{\bibinfo{volume}{85}},
  \bibinfo{pages}{115105} (\bibinfo{year}{2012}).

\bibitem[{\citenamefont{Chiu and Schnyder}(2014)}]{Chiu2014}
\bibinfo{author}{\bibfnamefont{C.-K.} \bibnamefont{Chiu}} \bibnamefont{and}
  \bibinfo{author}{\bibfnamefont{A.~P.} \bibnamefont{Schnyder}},
  \bibinfo{journal}{Phys. Rev. B} \textbf{\bibinfo{volume}{90}},
  \bibinfo{pages}{205136} (\bibinfo{year}{2014}).

\bibitem[{\citenamefont{Phillips and Aji}(2014)}]{Phillips2014}
\bibinfo{author}{\bibfnamefont{M.}~\bibnamefont{Phillips}} \bibnamefont{and}
  \bibinfo{author}{\bibfnamefont{V.}~\bibnamefont{Aji}},
  \bibinfo{journal}{Phys. Rev. B} \textbf{\bibinfo{volume}{90}},
  \bibinfo{pages}{115111} (\bibinfo{year}{2014}).

\bibitem[{\citenamefont{Chen et~al.}(2015)\citenamefont{Chen, Lu, and
  Kee}}]{Chen2015}
\bibinfo{author}{\bibfnamefont{Y.}~\bibnamefont{Chen}},
  \bibinfo{author}{\bibfnamefont{Y.-M.} \bibnamefont{Lu}}, \bibnamefont{and}
  \bibinfo{author}{\bibfnamefont{H.-Y.} \bibnamefont{Kee}},
  \bibinfo{journal}{Nature Communications} \textbf{\bibinfo{volume}{6}}
  (\bibinfo{year}{2015}).

\bibitem[{\citenamefont{Mullen et~al.}(2015)\citenamefont{Mullen, Uchoa, and
  Glatzhofer}}]{Mullen2015}
\bibinfo{author}{\bibfnamefont{K.}~\bibnamefont{Mullen}},
  \bibinfo{author}{\bibfnamefont{B.}~\bibnamefont{Uchoa}}, \bibnamefont{and}
  \bibinfo{author}{\bibfnamefont{D.~T.} \bibnamefont{Glatzhofer}},
  \bibinfo{journal}{Phys. Rev. Lett.} \textbf{\bibinfo{volume}{115}},
  \bibinfo{pages}{026403} (\bibinfo{year}{2015}).

\bibitem[{\citenamefont{Weng et~al.}(2014)\citenamefont{Weng, Liang, Xu, Rui,
  Fang, Dai, and Kawazoe}}]{Weng2014}
\bibinfo{author}{\bibfnamefont{H.}~\bibnamefont{Weng}},
  \bibinfo{author}{\bibfnamefont{Y.}~\bibnamefont{Liang}},
  \bibinfo{author}{\bibfnamefont{Q.}~\bibnamefont{Xu}},
  \bibinfo{author}{\bibfnamefont{Y.}~\bibnamefont{Rui}},
  \bibinfo{author}{\bibfnamefont{Z.}~\bibnamefont{Fang}},
  \bibinfo{author}{\bibfnamefont{X.}~\bibnamefont{Dai}}, \bibnamefont{and}
  \bibinfo{author}{\bibfnamefont{Y.}~\bibnamefont{Kawazoe}},
  \bibinfo{journal}{arXiv:1411.2175}  (\bibinfo{year}{2014}).

\bibitem[{\citenamefont{Xie et~al.}(2015)\citenamefont{Xie, Schoop, Seibel,
  Gibson, Xie, and Cava}}]{Xie2015}
\bibinfo{author}{\bibfnamefont{L.~S.} \bibnamefont{Xie}},
  \bibinfo{author}{\bibfnamefont{L.~M.} \bibnamefont{Schoop}},
  \bibinfo{author}{\bibfnamefont{E.~M.} \bibnamefont{Seibel}},
  \bibinfo{author}{\bibfnamefont{Q.~D.} \bibnamefont{Gibson}},
  \bibinfo{author}{\bibfnamefont{W.}~\bibnamefont{Xie}}, \bibnamefont{and}
  \bibinfo{author}{\bibfnamefont{R.~J.} \bibnamefont{Cava}},
  \bibinfo{journal}{arXiv:1504.01731}  (\bibinfo{year}{2015}).

\bibitem[{\citenamefont{Kim et~al.}(2015)\citenamefont{Kim, Wieder, Kane, and
  Rappe}}]{Kim2015}
\bibinfo{author}{\bibfnamefont{Y.}~\bibnamefont{Kim}},
  \bibinfo{author}{\bibfnamefont{B.~J.} \bibnamefont{Wieder}},
  \bibinfo{author}{\bibfnamefont{C.~L.} \bibnamefont{Kane}}, \bibnamefont{and}
  \bibinfo{author}{\bibfnamefont{A.~M.} \bibnamefont{Rappe}},
  \bibinfo{journal}{arXiv:1504.03807}  (\bibinfo{year}{2015}).

\bibitem[{\citenamefont{Yu et~al.}(2015)\citenamefont{Yu, Weng, Fang, Dai, and
  Hu}}]{Yu2015}
\bibinfo{author}{\bibfnamefont{R.}~\bibnamefont{Yu}},
  \bibinfo{author}{\bibfnamefont{H.}~\bibnamefont{Weng}},
  \bibinfo{author}{\bibfnamefont{Z.}~\bibnamefont{Fang}},
  \bibinfo{author}{\bibfnamefont{X.}~\bibnamefont{Dai}}, \bibnamefont{and}
  \bibinfo{author}{\bibfnamefont{X.}~\bibnamefont{Hu}},
  \bibinfo{journal}{arXiv:1504.04577}  (\bibinfo{year}{2015}).

\bibitem[{\citenamefont{Rhim and Kim}(2015)}]{Rhim2015}
\bibinfo{author}{\bibfnamefont{J.-W.} \bibnamefont{Rhim}} \bibnamefont{and}
  \bibinfo{author}{\bibfnamefont{Y.~B.} \bibnamefont{Kim}},
  \bibinfo{journal}{arXiv:1504.07641}  (\bibinfo{year}{2015}).

\bibitem[{\citenamefont{Chiu et~al.}(2015)\citenamefont{Chiu, Teo, Schnyder,
  and Ryu}}]{Chiu2015}
\bibinfo{author}{\bibfnamefont{C.-K.} \bibnamefont{Chiu}},
  \bibinfo{author}{\bibfnamefont{J.~C.} \bibnamefont{Teo}},
  \bibinfo{author}{\bibfnamefont{A.~P.} \bibnamefont{Schnyder}},
  \bibnamefont{and} \bibinfo{author}{\bibfnamefont{S.}~\bibnamefont{Ryu}},
  \bibinfo{journal}{arXiv:1505.03535}  (\bibinfo{year}{2015}).

\bibitem[{\citenamefont{Bian et~al.}(2015)\citenamefont{Bian, Chang, Sankar,
  Xu, Zheng, Neupert, Chiu, Huang, Chang, Belopolski et~al.}}]{Bian2015}
\bibinfo{author}{\bibfnamefont{G.}~\bibnamefont{Bian}},
  \bibinfo{author}{\bibfnamefont{T.-R.} \bibnamefont{Chang}},
  \bibinfo{author}{\bibfnamefont{R.}~\bibnamefont{Sankar}},
  \bibinfo{author}{\bibfnamefont{S.-Y.} \bibnamefont{Xu}},
  \bibinfo{author}{\bibfnamefont{H.}~\bibnamefont{Zheng}},
  \bibinfo{author}{\bibfnamefont{T.}~\bibnamefont{Neupert}},
  \bibinfo{author}{\bibfnamefont{C.-K.} \bibnamefont{Chiu}},
  \bibinfo{author}{\bibfnamefont{S.-M.} \bibnamefont{Huang}},
  \bibinfo{author}{\bibfnamefont{G.}~\bibnamefont{Chang}},
  \bibinfo{author}{\bibfnamefont{I.}~\bibnamefont{Belopolski}},
  \bibnamefont{et~al.}, \bibinfo{journal}{arXiv:1505.03069}
  (\bibinfo{year}{2015}).

\bibitem[{\citenamefont{Huh et~al.}(2015)\citenamefont{Huh, Moon, and
  Kim}}]{Huh2015}
\bibinfo{author}{\bibfnamefont{Y.}~\bibnamefont{Huh}},
  \bibinfo{author}{\bibfnamefont{E.-G.} \bibnamefont{Moon}}, \bibnamefont{and}
  \bibinfo{author}{\bibfnamefont{Y.~B.} \bibnamefont{Kim}},
  \bibinfo{journal}{arXiv:1506.05105}  (\bibinfo{year}{2015}).

\bibitem[{\citenamefont{Hatcher}(2002)}]{Hatcher2002}
\bibinfo{author}{\bibfnamefont{A.}~\bibnamefont{Hatcher}},
  \emph{\bibinfo{title}{Algebraic Topology}} (\bibinfo{publisher}{Cambridge
  University Press}, \bibinfo{year}{2002}).

\bibitem[{\citenamefont{Young and Kane}(2015)}]{Young2015}
\bibinfo{author}{\bibfnamefont{S.~M.} \bibnamefont{Young}} \bibnamefont{and}
  \bibinfo{author}{\bibfnamefont{C.~L.} \bibnamefont{Kane}},
  \bibinfo{journal}{arXiv:1504.07977}  (\bibinfo{year}{2015}).

\bibitem[{\citenamefont{Parameswaran et~al.}(2013)\citenamefont{Parameswaran,
  Turner, Arovas, and Vishwanath}}]{Parameswaran2013}
\bibinfo{author}{\bibfnamefont{S.~A.} \bibnamefont{Parameswaran}},
  \bibinfo{author}{\bibfnamefont{A.~M.} \bibnamefont{Turner}},
  \bibinfo{author}{\bibfnamefont{D.~P.} \bibnamefont{Arovas}},
  \bibnamefont{and}
  \bibinfo{author}{\bibfnamefont{A.}~\bibnamefont{Vishwanath}},
  \bibinfo{journal}{Nature Physics} \textbf{\bibinfo{volume}{9}},
  \bibinfo{pages}{299} (\bibinfo{year}{2013}).

\bibitem[{\citenamefont{Liu et~al.}(2014{\natexlab{b}})\citenamefont{Liu,
  Zhang, and VanLeeuwen}}]{Liu2014b}
\bibinfo{author}{\bibfnamefont{C.-X.} \bibnamefont{Liu}},
  \bibinfo{author}{\bibfnamefont{R.-X.} \bibnamefont{Zhang}}, \bibnamefont{and}
  \bibinfo{author}{\bibfnamefont{B.~K.} \bibnamefont{VanLeeuwen}},
  \bibinfo{journal}{Phys. Rev. B} \textbf{\bibinfo{volume}{90}},
  \bibinfo{pages}{085304} (\bibinfo{year}{2014}{\natexlab{b}}).

\bibitem[{\citenamefont{Fang and Fu}(2015)}]{Fang2015}
\bibinfo{author}{\bibfnamefont{C.}~\bibnamefont{Fang}} \bibnamefont{and}
  \bibinfo{author}{\bibfnamefont{L.}~\bibnamefont{Fu}}, \bibinfo{journal}{Phys.
  Rev. B} \textbf{\bibinfo{volume}{91}}, \bibinfo{pages}{161105}
  (\bibinfo{year}{2015}).

\bibitem[{\citenamefont{Shiozaki et~al.}(2015)\citenamefont{Shiozaki, Sato, and
  Gomi}}]{Shiozaki2015}
\bibinfo{author}{\bibfnamefont{K.}~\bibnamefont{Shiozaki}},
  \bibinfo{author}{\bibfnamefont{M.}~\bibnamefont{Sato}}, \bibnamefont{and}
  \bibinfo{author}{\bibfnamefont{K.}~\bibnamefont{Gomi}},
  \bibinfo{journal}{Phys. Rev. B} \textbf{\bibinfo{volume}{91}},
  \bibinfo{pages}{155120} (\bibinfo{year}{2015}).

\bibitem[{\citenamefont{Watanabe et~al.}(2015)\citenamefont{Watanabe, Po,
  Vishwanath, and Zaletel}}]{Watanabe2015}
\bibinfo{author}{\bibfnamefont{H.}~\bibnamefont{Watanabe}},
  \bibinfo{author}{\bibfnamefont{H.~C.} \bibnamefont{Po}},
  \bibinfo{author}{\bibfnamefont{A.}~\bibnamefont{Vishwanath}},
  \bibnamefont{and} \bibinfo{author}{\bibfnamefont{M.~P.}
  \bibnamefont{Zaletel}}, \bibinfo{journal}{arXiv:1505.04193}
  (\bibinfo{year}{2015}).

\bibitem[{\citenamefont{Bradley and Cracknell}(2010)}]{Bradley2010}
\bibinfo{author}{\bibfnamefont{C.}~\bibnamefont{Bradley}} \bibnamefont{and}
  \bibinfo{author}{\bibfnamefont{A.}~\bibnamefont{Cracknell}},
  \emph{\bibinfo{title}{The Mathematical Theory of Symmetry in Solids:
  Representation Theory for Point Groups and Space Groups}}
  (\bibinfo{publisher}{Oxford University Press}, \bibinfo{year}{2010}).

\bibitem[{\citenamefont{Liu et~al.}(2015)\citenamefont{Liu, Kriegner, Horak,
  Puggioni, Serrao, Yi, Frontera, Holy, Vishwanath, Rondinelli
  et~al.}}]{Liu2015}
\bibinfo{author}{\bibfnamefont{J.}~\bibnamefont{Liu}},
  \bibinfo{author}{\bibfnamefont{D.}~\bibnamefont{Kriegner}},
  \bibinfo{author}{\bibfnamefont{L.}~\bibnamefont{Horak}},
  \bibinfo{author}{\bibfnamefont{D.}~\bibnamefont{Puggioni}},
  \bibinfo{author}{\bibfnamefont{C.~R.} \bibnamefont{Serrao}},
  \bibinfo{author}{\bibfnamefont{D.}~\bibnamefont{Yi}},
  \bibinfo{author}{\bibfnamefont{C.}~\bibnamefont{Frontera}},
  \bibinfo{author}{\bibfnamefont{V.}~\bibnamefont{Holy}},
  \bibinfo{author}{\bibfnamefont{A.}~\bibnamefont{Vishwanath}},
  \bibinfo{author}{\bibfnamefont{J.~M.} \bibnamefont{Rondinelli}},
  \bibnamefont{et~al.}, \bibinfo{journal}{unpublished}  (\bibinfo{year}{2015}).

\end{thebibliography}
\end{document}